# Evidence for two distinct scales of current flow in polycrystalline Sm and Nd iron oxypnictides

A. Yamamoto, A. A. Polyanskii, J. Jiang, F. Kametani, C. Tarantini, F. Hunte, J. Jaroszynski, E. E. Hellstrom, P. J. Lee, A. Gurevich, D. C. Larbalestier
*National High Magnetic Field Laboratory, Florida State University, Tallahassee, FL 32310, USA*
Z. A. Ren, J. Yang, X. L. Dong, W. Lu, Z. X. Zhao
*National Laboratory for Superconductivity, Institute of Physics and Beijing National Laboratory for Condensed Matter Physics, Chinese Academy of Sciences, P.O. Box 603, Beijing 100190, P.R. China*

**Abstract**
Early studies have found quasi-reversible magnetization curves in polycrystalline bulk rare-earth iron oxypnictides that suggest either wide-spread obstacles to intergranular current or very weak vortex pinning. In the present study of polycrystalline samarium and neodymium rare-earth iron oxypnictide samples made by high pressure synthesis, the hysteretic magnetization is significantly enhanced. Magneto optical imaging and study of the field dependence of the remanent magnetization as a function of particle size both show that global currents over the whole sample do exist but that the intergranular and intragranular current densities have distinctively different temperature dependences and differ in magnitude by about 1000. Assuming that the highest current density loops are restricted to circulation only within grains leads to values of $\sim 5 \times 10^6$ A/cm$^2$ at 5 K and self field, while whole-sample current densities, though two orders of magnitude lower are 1000-10000 A/cm$^2$, some two orders of magnitude higher than in random polycrystalline cuprates. We cannot yet be certain whether this large difference in global and intragrain current density is intrinsic to the oxypnictides or due to extrinsic barriers to current flow, because the samples contain significant second phase, some of which wets the grain boundaries and produces evidences of SNS proximity effect in the whole sample critical current.

**Introduction:**

The recent discovery of superconductivity in the LaFeAsO$_{1-x}$F$_x$ compound [1] has stimulated a rapid exploration of superconductivity in the rare earth iron oxypnictides [2-14]. It has now been established that the iron oxypnictides can be superconducting when doped to $x$ ~0.05-0.2 and that they can have transition temperature $T_c$ above 40 K when La is replaced by Ce [5] and above 50 K by Pr, Nd, Sm and Gd [7-11]. In a recent paper [12] we addressed the issue of electromagnetic granularity in polycrystalline La iron oxypnictides, finding an asymmetric $M(H)$ loop that indicated an irreversible moment due to hysteretic bulk currents that was almost as small as the reversible magnetization of the superconducting state. In that case we were not able to distinguish definitively between a state where the intragrain pinning was very weak, leading to very low intragrain current densities or to the state where currents were largely confined to the intragrain regions and might have been rather high. Based on the rather high upper critical field $B_{c2}(0)$ values of 63-65 T observed by Hunte *et al.* [6] on the same sample and the nanoscale coherence length $\xi$ ($\xi^{ab}(0)$ = 5 nm, $\xi^c(0)$ = 1.2 nm), we would expect strong vortex pinning even from naturally occurring atomic-scale defects. By analogy to randomly oriented polycrystalline cuprates which also show small hysteretic current loops and large intragrain current densities of $10^5$-$10^6$ A/cm$^2$ at 4 K [15,16], we proposed [12] that electromagnetic granularity was likely to be characteristic of polycrystalline oxypnictides. Evidence for granular behavior were also presented in two subsequent studies of a Sm oxypnictide (Senatore *et al.* [13]) and a Nd oxypnictide (Prozorov *et al.* [14]), the latter of which also presented magneto optical images of confined current flow within intragranular regions. In the present study, we extend our initial examination of the length scales of current flow in La oxypnictide by combining magneto optical imaging, remanent magnetic field analysis and powdering of the sample to conclusively demonstrate the presence of two distinctly different length scales of current flow in our dense Nd and Sm oxypnictides and to show that the temperature dependence of the inter- and intra-granular current densities are quite different and that their ratio is strongly dependent on temperature.

**Experimental details:**

The polycrystalline SmFeAsO$_{0.85}$ and NdFeAsO$_{0.94}$F$_{0.06}$ bulk samples were synthesized by solid state reaction under a high pressure [8,10]. SmAs (or NdAs, pre-sintered) powder and Fe, Fe$_2$O$_3$, FeF$_2$ powders were mixed together according to the nominal stoichiometric ratio then ground thoroughly and pressed into small pellets. The pellets were sealed in boron nitride crucibles and sintered in a high pressure synthesis apparatus under the pressure of 6 GPa at 1250°C for 2 hours.

Microstructural observations were performed using field emission scanning electron microscope (Carl Zeiss 1540 ESB and XB) and a laser scanning confocal microscope (Olympus OLS3100). Resistivity measurements were performed by the conventional four-point-probe method using a Quantum Design PPMS. Magnetization of the samples was measured by a SQUID magnetometer (Quantum Design: MPMS-XL5s) and a 14 T vibrating sample magnetometer (Oxford) with field parallel to the broad face. Magneto Optical imaging with a 5 μm thick Bi-doped iron-garnet indicator film placed directly onto the sample surface was used to image the normal field component $B_z$ produced by magnetization currents induced by solenoidal fields of up to 0.12 T applied perpendicular to the imaged surface [17,18].

**Results:**

The temperature dependences of resistivity for the SmFeAsO$_{0.85}$ and NdFeAsO$_{0.94}$F$_{0.06}$ bulk samples are shown in Fig. 1. Resistivity began to drop at 57 and 51 K and vanished below 51 and 44 K for the Sm and Nd samples, respectively. The calculated resistivity at 300 K for the Sm and Nd samples were 2.3 and 2.0 mΩcm and the $RRR = \rho(300\ K)/\rho(60\ K)$ were 3.7 and 3.4, respectively. By contrast the $RRR$ value of 17 was observed for the earlier studied LaFeAsO$_{0.89}$F$_{0.11}$ bulk sample [4,12], which might suggest that the Sm and Nd samples are less strongly doped and that the actual doping state may not be well represented by the nominal composition.

Figure 2 shows the temperature dependences of magnetization in zero-field-cooled (ZFC) and field-cooled (FC) states under an external field of 1 mT. Onset magnetic $T_c$ values were found to be 53 K for the Sm sample and 49 K for the Nd sample. Compared to the LaFeAsO$_{0.89}$F$_{0.11}$ sample [12] and other reported samples [13,14], superconducting transitions are rather sharp ($\Delta T_c \sim 10$ K), indicative of bulk scale shielding currents flowing in both samples, even close to $T_c$. Within the uncertainty limits produced by the demagnetization fields of these imperfectly shaped samples, we conclude that the shielded volumes were 100%.

Figure 3(a) shows magnetic hysteresis loops at 5, 20, 30 and 40 K obtained by VSM for the SmFeAsO$_{0.85}$ bulk sample. So far very small hysteresis loops were reported for polycrystalline iron oxypnictides [12,13]. However, this Sm sample shows quite large hysteresis loops, which implies either strong flux pinning and/or good intergranular coupling. Slightly smaller hysteresis loop widths were observed in the Nd sample. Similar to the previously studied LaFeAsO$_{0.89}$F$_{0.11}$ sample [12], a paramagnetic background was observed to all curves taken below $T_c$, which can be well fit by a Langevin expression [19].

Figure 3(b) shows the magnetic field dependence of the critical current density $J_c$ derived from the hysteresis loop width using the extended Bean model $J_c = 20\ \Delta m/Va(1-a/3b)$ for the Sm bulk sample taking the full sample dimensions of 2×1×0.6 mm$^3$. This expression yields a $J_c$ of 10,000-30,000 A/cm$^2$ at 5 K, which is nearly independent of field over the range of 4-14 T. $J_c$ for the Nd bulk sample is lower, as is shown in Fig. 3(b). Broad maximum in $J_c(B)$ was observed at 20, 30 and 40 K, a result also noted in ref [13]. Since the contribution of currents circulating on smaller length scales to the hysteresis loop is large, as discussed later, the critical current density shown in Fig. 3(b) is likely to be overestimated, since our later, size-dependent studies allow us to deduce that the contribution of global currents to the hysteretic magnetization is less than that produced by the locally circulating currents. The values of $J_c$ of ~10,000 A/cm$^2$ at 5 K in self field are about two orders of magnitude smaller than those of the practical low temperature superconductors, MgB$_2$ and textured cuprate high temperature superconductors where high $J_c$ of $10^6 - 10^7$ A/cm$^2$ are obtained. However, these $J_c$ values are distinctly better than randomly oriented polycrystalline cuprates, which are typically ~100 A/cm$^2$ [16]. It is also important to note that the relatively large hysteresis loop and finite $J_c$ values observed in the Sm and Nd samples are strong contrast to the almost reversible magnetization observed in the LaFeAsO$_{0.89}$F$_{0.11}$ sample where bulk flux shielding was not visible in magneto optical images.

Figure 4(a) shows a whole sample image of the SmFeAsO$_{0.85}$ bulk obtained by scanning electron microscopy. The bulk sample has a relative density of ~90%. High magnification images of the polished surface of the SmFeAsO$_{0.85}$ reveals plate-like grains of the superconducting phase in a size of ~20 µm, as shown in Fig. 4(b). Atomic number sensitive back-scattered electron (BSE) imaging also revealed that the sample contains multiple impurity phases, with the two most prominent phases, as shown in Fig. 4(b), identified by EDS as Sm$_2$O$_3$ (white contrast) and Fe-As glassy phase (dark contrast). It does appear that the microstructure is homogeneous on the macro-scale, although the Nd bulk sample showed macroscale inhomogeneity on a scale of several hundreds micrometers (see Fig. 8(a)). The larger fraction of impurity phases in the Nd bulk is consistent with its lower $J_c$ compared to that of the Sm bulk.

Figure 4(c) shows a laser scanning microscope image of the crushed pieces of SmFeAsO$_{0.85}$ with particle size of ~500 µm, while Figure 4(d) shows a scanning electron microscope image of the ground powder. Most of the particles are less than 100 µm in size, having an average size of 20-50 µm. Higher magnification SEM images revealed that each particle still contains several to several tens of grains, even after fine grinding.

In order to make a more explicit test of the scale over which currents flow, we made remanent magnetization analysis on polycrystalline samples of different size. Both intact bulk samples were exposed to many cycles of ever increasing magnetic field $H_a$, followed by removal of the field and measurement of the remanent moment, $m_R$ [12,20-22]. For a pure and homogeneous sample, we expect flux to penetrate when $H_a/(1-D)$ first exceeds the lower critical field $B_{c1}$, where $D$ is the relevant demagnetizing factor. For weakly coupled polycrystals, where the weakness could occur either at grain boundaries or at non-superconducting second phases, flux penetration occurs locally at lower fields than into the grains. Therefore information about the size of current loops derived from $m_R$ (which is proportional to the product of $J_c$ and current loop size) can be extracted from the dependence of $m_R$ on the applied field $H_a$ and the particle size. Thus, after measurement of each whole sample, we then crushed them into tens of pieces and remeasured $m_R$, finally gently grinding the crushed pieces to powder in a mortar before remeasuring $m_R$. $T_c$ did not change by crushing or powdering.

Figure 5(a) shows the remanent magnetization as a function of increasing applied field for the SmFeAsO$_{0.85}$ intact large bulk, a second, smaller bulk, the crushed pieces and ground powder. The $m_R$ data are normalized by their respective sample masses of 30.0, 6.0, 14.2 and 4.3 mg. For the bulk samples, remanent magnetization began to increase on increasing the applied field above ~5 mT, consistent with the reported $B_{c1}$ of 5 mT for LaFeAsOF [23], and a shoulder appeared at ~80 mT in the $m_R$ transition. This behavior is more clear in the derivative of $m_R(H_a)$ shown in Figure 5(b) where two quite separate low-field peaks appear at 35 mT for the larger and at 18 mT for the smaller of the two bulk samples. By contrast, the second, higher field peak appeared at the same field of 150 mT for both samples. That the first peak strongly depends on sample size indicates that the bulk current loop is also size dependent, because the field of first penetration should be proportional to $H_p \sim J_c^{global} \times$ (sample size). This conclusion is strengthened by further suppression of the first peak in the crushed pieces and its disappearance almost to zero in the ground powder. In contrast, the second peak was found to be size independent, even after fine powdering, which means that the second peak is caused by locally circulating currents with current loop size less than the powder size of 20-50 μm.

Magneto Optical Imaging (MOI) was performed to directly observe the local magnetic flux structure of the intact samples. Figure 6 shows MO images on a well polished surface of the SmFeAsO$_{0.85}$ bulk with a thickness of 440 μm. The light microscope image in Fig. 6(a) does not reveal any macroscale defects such as cracks or connected second phase fields that would block whole-sample current flow. Under the zero-field-cooling (ZFC) condition we observe a bulk Meissner state when a field of 4 mT is applied at 6 K, as shown in Fig. 6(b), which indicates that the surface shielding current flows over the whole sample. As the external field increases, flux starts to penetrate at ~6 mT and first reaches the center of the sample at ~15 mT. However what the MO images clearly show is that flux penetration is quite inhomogeneous. As follows from Figs. 6(c) and (d), there are many 20-50 μm size black-appearing spots of strong flux shielding in the flux penetrated regions, indicative of local circulating currents with higher current density than the matrix. Similar electromagnetically granular behavior was also observed in the Nd bulk sample, as will be presented in Fig. 8. After the removal of the external field, the remanent trapped fields are shown in Figs. 6 (e) and (f). These trapped fields were more homogeneous after applying a lower external field of 40 mT and, in contrast, many higher $J_c$ (now white) spots appeared after applying 120 mT. These results are also consistent with the remanent magnetization data of Fig. 5(b), where the second peak appeared at 150 mT. Figures 6 (g) and (h) show MO images under $\mu_0 H_{ex}$ = 0 mT at $T$ = 6.4 and 20 K, respectively, for the sample field-cooled in $\mu_0 H_{ex}$ = 120 mT. It is particularly noteworthy that the strongly coupled local regions were clearly visible at 6.4 K, however that on raising the temperature to 20 K, a quasi-ideal, roof-top pattern then appears superimposed on the less visible granular structure. Therefore, the magnetic granularity is more pronounced below 15 K and becomes progressively less visible at higher temperatures. Field-cooled MO contrast of the roof-top pattern of the trapped flux state was visible up to 48 K, indicating persistent bulk currents up to almost 90% of $T_c \sim 53$ K.

This temperature-dependent granular behavior is quite prominent in flux density profiles. Figure 7 shows magnetic flux profiles $B(x)$ taken at 6.4 K (a) and 20 K (b) for the Sm sample field-cooled under $\mu_0 H_{ex}$ = 120 mT. The profiles are taken along the same line (line 1) as indicated by arrows in Fig. 6 (g) and (h), respectively. A bulk scale critical state with global $J_c$ of $1.9 \times 10^3$ A/cm$^2$ at 20 K was deduced from this flux gradient, but the gradients are indeed strongly perturbed at 6 K by the local domains, which correspond directly to the white spots in Fig. 6.

Figure 8(a) shows MO images taken on the polished surface of the NdFeAsO$_{0.94}$F$_{0.06}$ bulk. Topographic light microscopy images showed dark color impurity phases 100-300 μm in size, which were identified to be Nd$_2$O$_3$ particles and Fe-As glassy phase by EDS analyses. Figures 8 (b)-(d) show the MO images of the sample after zero-field cooling to 6 K. Fig. 8(b) shows that flux penetrates preferentially through the larger impurity phases and reaches the center of the sample under an external field of 40 mT. Granular behavior was also observed after applying a higher magnetic field of 120 mT, as shown in Fig. 8(e). The 10-20 μm size of the strongly connected white spots was smaller than those of the Sm sample, consistent with our observation that the ~5 μm grain size of the Nd sample is about half that of the Sm sample. Magnetic flux profiles $B_z(x)$ taken along the line 2 (b) in Fig. 8 (c) and (d) are shown in Fig. 9, where it is also seen that a bulk critical state is quite visible at 20 K, but that much more inhomogeneous behavior is observed at low-temperatures.

Figure 10 shows the derivative of $m_R(H_a)$ for the SmFeAsO$_{0.85}$ bulk at 5, 10, 15, 20, 30 and 40 K, while the inset figure shows the temperature dependence of the first and second peaks in the derivative of the $m_R(H_a)$ plots. The second high-field peak shows a strong temperature dependence and a marked shift towards higher fields below ~15 K, indicative either of a strongly increasing local critical current density or/and a growing current loop size. On the other hand, the low-field global-current peak showed a much weaker temperature dependence. These quite different temperature dependences result in a merging of the two peaks at high temperatures above 15 K, as is consistent with the MO images seen in Figure 6.

Figure 11 shows the derivative of $m_R(H_a)$ for the NdFeAsO$_{0.94}$F$_{0.06}$ bulk sample at 5, 10, 20, 30 and 35 K. Similar to the Sm sample, the second peak shifts rapidly as the temperature decreases. A third peak (~110 mT) appeared at 5 K. Since the ratio of the second and third peak positions is ~4, and this value is comparable to the square root of the resistivity anisotropy ratio ~15 [24], the peak split might be due to an anisotropic $J_c$. The inset to Figure 11 shows the temperature dependence of the first, second and third peaks. As for the Nd sample, the temperature dependence of second/third peaks showed strong temperature dependence below ~10 K. A minor peak was observed at ~300 mT and was almost temperature independent. This peak at 200-300 mT was also observed in the Sm and LaFeAsO$_{0.89}$F$_{0.11}$ samples. It might have its origin in a minority paramagnetic phase and is not further discussed in this paper.

Global and local $J_c$ values for the Sm and Nd bulk samples were calculated independently from the peak positions in the derivative of $m_R(H_a)$. For $J_c$ calculation we assumed a thin plate in parallel field for the bulk current and spherical grains for the local current. By neglecting the demagnetization factor and anisotropy of $J_c$, both of which are unknown at this stage of oxypnictide studies, the Bean model shows that the global and local $J_c$ can be given as $J_c^{global} = 2H_{peak1}/w$ and $J_c^{local} = 2.81\, H_{peak2}/r$, where $w$ is the thickness of a bulk sample and $r$ is the grain radius. From microstructural analyses, grain sizes were estimated to be 10 μm for the Sm sample and 5 μm for the Nd sample, leading to the global and local $J_c$ as a function of temperature shown in Fig. 12 and Fig. 13, respectively. Global $J_c$ values were also calculated from MO flux profiles and these values are also shown in Fig. 12, as is $J_c$ calculated from the width of the flux penetration front in the ZFC image using formulae in refs [17,18,25]. The agreement between these various methods of extracting the global $J_c$ is excellent. The temperature dependence of the global $J_c$ is almost linear except near $T_c$ and $J_c$ values at 5 K of 3900 and 2100 A/cm$^2$ for the Sm and Nd samples, respectively, are obtained.

The temperature dependence of the local $J_c$ is shown in Fig. 13, gives the values of $4.5\times10^6$ and $6.7\times10^6$ A/cm$^2$ at 5 K for the Sm and Nd samples are deduced, respectively. These values are considerably higher than those recently deduced by Prozorov et al. ($3\times10^5$ A/cm$^2$ [13]), by Senatore et al. ($6\times10^4$ A/cm$^2$ for 10 μm grain size and $6\times10^6$ A/cm$^2$ for 0.1 μm [14]) and by Wang et al. ($2\times10^5$ A/cm$^2$ [26]). Both Sm and Nd samples showed strong upward curvature in $J_c(T)$ below ~20 K.

Finally we plot the ratio of the global and local $J_c$ values in Fig. 14. The ratio $J_c^{global}(T)/J_c^{local}(T)$ increases as temperature decreases, reaches a maximum at ~25 K and drops rapidly at low temperatures. The ratio never exceeds 0.004 and falls to ~0.001 at ~5 K when the granular behavior revealed by the MO imaging and remanent field analysis becomes so marked.

**Discussion:**

The sample-size dependent remanent magnetization showed that two (or even three in the case of the Nd sample) distinct scales of current flow exist in these iron oxypnictide samples. The spatial variation of the scales is directly demonstrated by MO imaging where obvious spots of higher $J_c$ appear in a lower $J_c$ matrix in a field-cooled images. Here we discuss the origins of this electromagnetically granular behavior, especially the issue of whether the granularity is intrinsic or extrinsic to these samples.

First we discuss the scales of the global and local currents. The strong sample size dependence of the first peak of the remanent magnetization shown in Fig. 5(b) strongly suggests that a global current does circulate over the whole sample. This is in strong contrast to the previously reported data on a LaFeAsO$_{0.89}$F$_{0.11}$ sample where almost no bulk current was observed and only one, size-independent (the one equivalent to the second peak in the present data) peak appeared in the remanent magnetization [12]. Like the La sample data, the size independence of this second peak in Fig. 5(b) shows that the local current loop size is less than the powder size of ~50 μm shown in Fig. 4(d). Independent evidence of the scale of the locally high $J_c$ domains is provided by the white spots with size of 10-50 μm in the MO images shown in Fig. 6(g). We therefore conclude that a significant fraction of the local current loops are on the scale of the grain size, which lies principally in the 5-10 μm range.

An important scientific and technological issue is whether the substantial restriction of strong current density within the grains of this untextured, or very weakly textured sample (unpublished EBSD shows few signs of low angle grain boundaries in the samples [27]) is intrinsic to the oxypnictides or due to extrinsic features of these particular samples. So far as the possibility of intrinsic granularity is concerned, one important factor is likely to be the low carrier density of ~$10^{21}$ cm$^{-3}$ [4] which is very similar to that found in the cuprates and which produces an intrinsic weak-link behavior at high-angle cuprate grain boundaries [28,29]. However, many additional specific features of the cuprates, including their sensitivity to local oxygen concentration, proximity of the superconducting state to the Mott metal-insulator transition, and cation disorder [30], also play important roles in depressing the superconducting order parameter at grain boundaries. Explicit understanding of the oxypnictides awaits single grain boundary studies that are not yet possible, so for now we restrict our discussion to extrinsic factors that may be playing a role in the behavior of these two Nd and Sm samples, that actually show much better global current flow than the La oxypnictide sample first studied [12].

First it is clear from the MO images of Fig. 8(b) that the macroscale impurity phases (principally $RE_2O_3$ and the glassy Fe-As phase) shown in Figs. 4(b) and 8(a) do allow flux to easily penetrate the Nd sample. It is also seen that the global $J_c$ of the Nd sample is less than the Sm sample which has fewer second phases, a correlation that does suggest that macroscale impurities limit the global current, in analogy to the porous polycrystalline $MgB_2$ where voids and impurity insulating MgO significantly reduce percolating current path and limit the normal-state conductivity and critical current density [31,32].

Second it is possible that many superconducting grains are isolated from each other by non-superconducting layers, particularly the grain-boundary wetting, Fe-As amorphous phase. A TEM study in progress does show that some grain boundaries are covered by such wetting phases with a thickness of several tens of nanometers, although some clean grain boundaries are also observed [27]. Depending on the thickness and properties of the wetting phase, such layers might act as barriers that completely decouple the grains or act as weakly coupled Josephson junctions with rather different temperature- and field-dependent critical current density to the vortex pinning current density circulating within grains. Indeed the strong temperature dependence of $J_c^{global}(T)/J_c^{local}(T)$ in Figs. 12-14 shows very clearly that two different mechanisms control the inter- and intra-granular current densities. The temperature dependence of the global $J_c$ near $T_c$ with upward curvature is found to be well fitted by a quadratic function $J_c \propto (T_c - T)^2$ suggestive of an SNS proximity coupled Josephson junctions [33,34] which has also been observed on low-angle grain boundaries in cuprate superconductors [35,36]. Therefore we may conclude that the intergranullar current transport in the present samples is limited by the proximity coupled conductive wetting glassy As-Fe phase or by the intrinsic weak coupling. $H_{c2}$ measurements on these two samples suggest that $H_{c2}(0)$ is well over 100 T, making the coherence length shorter than 2 nm, thus adding weight to the concern that there is an intrinsic aspect to the limitation of current across high angle grain boundaries. The already noted negligible global current observed in the $LaFeAsO_{0.89}F_{0.11}$ sample surprised us due to its lower normal-state resistivity (~0.15 mΩcm) and higher *RRR* of 15 (~3 for the present samples), both factors which might suggest higher intergranular $J_c$ behavior for the $LaFeAsO_{0.89}F_{0.11}$ sample. However, the relatively low normal-state resistivity at $T_c$ (~0.5 mΩcm) of these samples considering an influence of current percolation induced by the second phases is not inconsistent with strongly suppressed $J_c^{global}(T)/J_c^{local}(T)$ by conductive wetting phases. We therefore conclude that our present data is insufficient to decide on the balance between intrinsic and extrinsic limitation of $J_c$ in the present samples and that there is evidence to support both mechanisms.

Another point to address is the comparative behavior of the polycrystalline oxypnictides to randomly oriented polycrystalline $MgB_2$ and cuprates. The global $J_c$ values of 1000-10000 A/cm$^2$ at 5 K obtained for the Sm and Nd samples is significantly lower than that seen in random bulks of $MgB_2$ which generally attain $10^6$ A/cm$^2$ at 4 K [37,38]. However, it was very early established that grain boundaries were not intrinsic obstacles to current flow in $MgB_2$ [39]. Of greater interest is the comparison to the least anisotropic of the cuprates, those with the RE-123 structure, randomly oriented polycrystalline examples of which have global $J_c$ values of only ~100 A/cm$^2$ at 4 K [15,16]. Thus it may be that any intrinsic weak-link effect at oxypnictide grain boundaries is less serious than in the cuprates. The lower $H_{c2}$ anisotropy $\gamma$ and higher carrier density in the iron oxypnictides ($\gamma$~4) as compared to the cuprates such as YBCO ($\gamma$~7) could be the reasons for the higher bulk $J_c$ in the randomly oriented Sm and Nd polycrystalline samples.

Finally, the temperature dependence of local $J_c$ is briefly discussed. A rather linear temperature dependence of local $J_c$ is observed near $T_c$ for both the Sm and Nd samples, as shown in the inset to the Fig. 13. This temperature dependence, very high $J_c(0) \sim 8 \times 10^6$ A/cm$^2$ and wide magnetization hysteresis loop indicate that strong intragrain vortex pinning is present in these samples. On the other hand, an upward curvature in $J_c(T)$ is observed below ~20 K and $J_c(T)$ decreases rapidly with increasing temperature. This behavior may indicate significant thermal fluctuation of vortices which result in a similar upturn of $J_c(T)$ at low-temperature in the cuprate single crystals [17]. The $J_c(T)$ curves at low-temperature can be fitted by an exponential function as shown in the inset to Fig. 13 using an equation $J_c = A (1 - T/T_c) \exp(-T/T_0)$, with $T_0$ = 8-10 K. The exponential factor can be understood using the following simple model. Vortex fluctuations smear out the pinning potential, reducing the elementary pinning forces by the Debye-Waller factor $\exp(-\langle u^2\rangle/a^2)$ where $u$ is the thermal displacement of a vortex from the pinning well of size $a$ [40]. Then from the equipartition theorem $G\langle u^2\rangle = k_B T$, we obtain $\exp(-\langle u^2\rangle/a^2) = \exp(-T/T_0)$. Here the depinning temperature $T_0 = Ga^2/k_B$, depends on the size of pinning centers $a$, and the Campbell pinning spring constant $G$. A crossover of the linear and exponential behaviors occurred at ~20 K as shown in the inset to the Fig. 13, indicating that thermal fluctuations result in the rapid decrease of the local $J_c$ at high temperatures. The deduced depinning temperature is $T_0$ = 8.8 K turned out to be rather low as compared to $T_c$ = 53 K, indicating either that the size of vortex pinning centers is small or the spring constant is small perhaps due to strong anisotropy of these sample. Noteworthy too is the fact that though the nominal doping level of the SmFeAsO$_{0.85}$ ($x$ = 0.15 oxygen deficiency) and NdFeAsO$_{0.94}$F$_{0.06}$ ($x$ = 0.06 fluorine doping) samples are quite different, the deduced temperature dependence of the local $J_c$ is quite similar, indicating either that the actual doping states differ from the nominal and/or that vortex pinning is less sensitive to RE type and doping level than in the cuprates.

**Summary:**

Two distinct scales of current in the polycrystalline SmFeAsO$_{0.85}$ and NdFeAsO$_{0.94}$F$_{0.06}$ samples were quantitatively evaluated by remanent magnetization measurements and local magneto optical imaging. A strong sample-size dependent remanent magnetization characteristic was found, however independent MO observations of the Bean critical state over the whole sample showed that a global current can exist in iron oxypnictides with $J_c$ of the order of ~4000 A/cm$^2$ at 5 K and self field. The temperature dependence of the global and local $J_c$ was independently derived from the penetration fields obtained from the global magnetization measurements and it was found that the two currents had quite different temperature dependences and that their ratio $J_c^{global}(T)/J_c^{local}(T)$ never exceeded ~0.004. An SNS proximity coupled intergranullar current was suggested from the temperature dependence of global $J_c$ near $T_c$. The local $J_c$ value at self field and 5 K is estimated to be ~$5\times10^6$ A/cm$^2$. Temperature dependence of local $J_c$ showed strong upturn below 15 K which is similar to the cuprate superconductors. Similar to the cuprates, we also found evidence for electromagnetic granularity, though with not quite so much reduction of $J_c$ in polycrystalline bulks. To fully understand the balance of intrinsic and extrinsic factors, study of individual grain boundaries of known misorientations will be very helpful.

**Acknowledgement:**

Work at the NHMFL was supported by IHRP 227000-520-003597-5063 under NSF Cooperative Agreement DMR-0084173, by the State of Florida, by the DOE, by the NSF Focused Research Group on Magnesium Diboride (FRG) DMR-0514592 and by AFOSR under grant FA9550-06-1-0474. One of the authors (AY) is supported by a fellowship of the Japan Society for the Promotion of Science.

**Figure captions:**

**Figure 1** Temperature dependence of resistivity for the SmFeAsO$_{0.85}$ and NdFeAsO$_{0.94}$F$_{0.06}$ bulk samples. Inset shows resistivity near $T_c$.

**Figure 2** Temperature dependence of magnetization under zero-field cooling (ZFC) and field-cooling (FC) conditions in an external field of 1 mT for the SmFeAsO$_{0.85}$ and NdFeAsO$_{0.94}$F$_{0.06}$ bulk samples.

**Figure 3** Magnetization hysteresis loops (a) and magnetic field dependence of $J_c$ (b) at 5, 20, 30 and 40 K for the bulk SmFeAsO$_{0.85}$. $J_c$ data for NdFeAsO$_{0.94}$F$_{0.06}$ bulk at 5 K (dashed line) is also shown for comparison.

**Figure 4** Scanning electron microscopy (SEM) images for the polished surface of the SmFeAsO$_{0.85}$ bulk sample (a, b). Confocal Laser Scanning Microscope image for the crushed pieces (c) and SEM image for the ground powder for the SmFeAsO$_{0.85}$.

**Figure 5** (a) Remanent magnetization ($m_R$) as a function of the maximum applied field at 5 K for the SmFeAsO$_{0.85}$ intact large bulk, small bulk, crushed pieces and ground powder. The data are normalized by the sample mass, 30.0, 6.0, 14.2 and 4.3 mg, respectively. (b) Derivatives of $m_R$ at 5 K.

**Figure 6** Magneto-optical images on the SmFeAsO$_{0.85}$ bulk sample. (a) Light microscopy image of a polished surface for the SmFeAsO$_{0.85}$ bulk. (b)-(f) MO images of different stages of magnetic flux penetration into the sample for ZFC at $T = 6$ K and $\mu_0 H_{ex} = 4$ mT (b), $\mu_0 H_{ex} = 12$ mT (c), $\mu_0 H_{ex} = 120$ mT (d), $\mu_0 H_{ex} = 0$ mT after applying 40 mT (e), $\mu_0 H_{ex} = 0$ mT after applying 120 mT (f). (g), (h) MO images under $\mu_0 H_{ex} = 0$ mT at $T = 6.4$ K (g) and 20 K (h) for the sample field-cooled (FC) in $\mu_0 H_{ex} = 120$ mT. Arrows shown in (g) and (h) indicate lines 1 and 2 for magnetic flux profiles as shown in Fig. 7.

**Figure 7** Magnetic flux profiles $B_z(x)$ taken at 6.4 K (a) and 20 K (b) along the line 1 in Fig. 6 (g) and (h), respectively. The sample was field-cooled under $\mu_0 H_{ex} = 120$ mT down to 6 K followed by a removal of external field and then temperature was increased.

**Figure 8** Magneto-optical images on the polished surface of the NdFeAsO$_{0.94}$F$_{0.06}$ bulk sample. (a) Surface image by the optical microscopy. MO images of different stages of magnetic flux penetration into the sample for ZFC at $T = 6$ K and $\mu_0 H_{ex} = 40$ mT (b), $\mu_0 H_{ex} = 0$ mT after applying 40 mT (c), $\mu_0 H_{ex} = 120$ mT (d), $\mu_0 H_{ex} = 0$ mT after applying 120 mT (e).

**Figure 9** Magnetic flux profiles $B_z(x)$ taken along the line 2 in Fig. 8 (a) at 6.2 K (a) and 19 K (b). The sample was field-cooled under $\mu_0 H_{ex} = 120$ mT down to 6 K followed by removal of the external field and increase of temperature to 19 K.

**Figure 10** Derivatives of remanent magnetization as a function of the maximum applied field at 5, 10, 15, 20, 30 and 40 K for the SmFeAsO$_{0.85}$ small bulk. Data are normalized by sample mass. Inset shows the temperature dependences of the peak fields $\mu_0H_{peak}$ for the first and second peaks.

**Figure 11** Derivatives of remanent magnetization as a function of the maximum applied field at 5, 10, 20, 30 and 35 K for the NdFeAsO$_{0.94}$F$_{0.06}$ bulk sample. Data are normalized by sample mass. Inset shows the temperature dependences of the peak fields $\mu_0H_{peak}$ for the first, second and third peaks.

**Figure 12** Temperature dependence of global critical current density $J_c^{global}(T)$ for the polycrystalline SmFeAsO$_{0.85}$ and NdFeAsO$_{0.94}$F$_{0.06}$ bulk samples obtained from the remanent magnetization analysis (filled), magneto optical $B(x)$ flux profile analysis for the FC conditions and the width of the flux penetration front in the ZFC conditions using Brandt's expression [21] for a strip in perpendicular field.

**Figure 13** Temperature dependence of critical current density of locally circulating current $J_c^{local}(T)$ for the polycrystalline SmFeAsO$_{0.85}$ and NdFeAsO$_{0.94}$F$_{0.06}$ bulk samples obtained from remanent magnetization analysis. Inset shows log-scale plots for the SmFeAsO$_{0.85}$ experimental data with an exponential fitting and linear fitting.

**Figure 14** Temperature dependence of the ratio $J_c^{global}(T)/J_c^{local}(T)$ for the polycrystalline SmFeAsO$_{0.85}$ and NdFeAsO$_{0.94}$F$_{0.06}$ bulk samples.

Figure 1

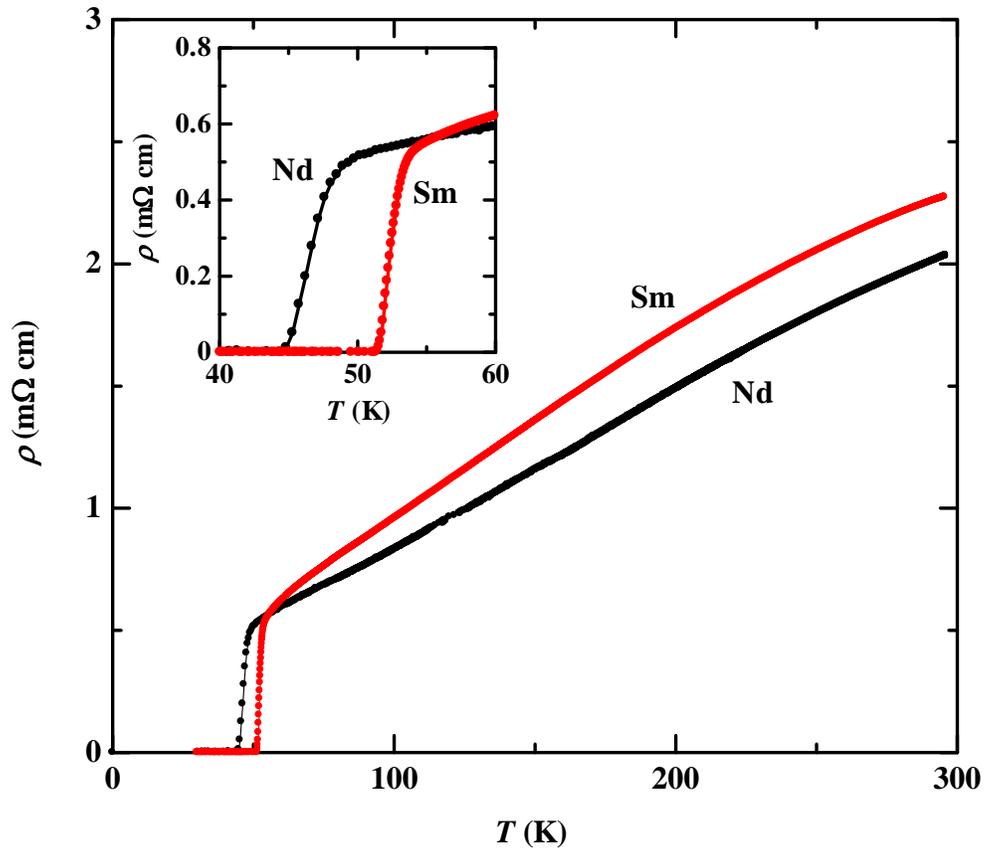

Figure 2

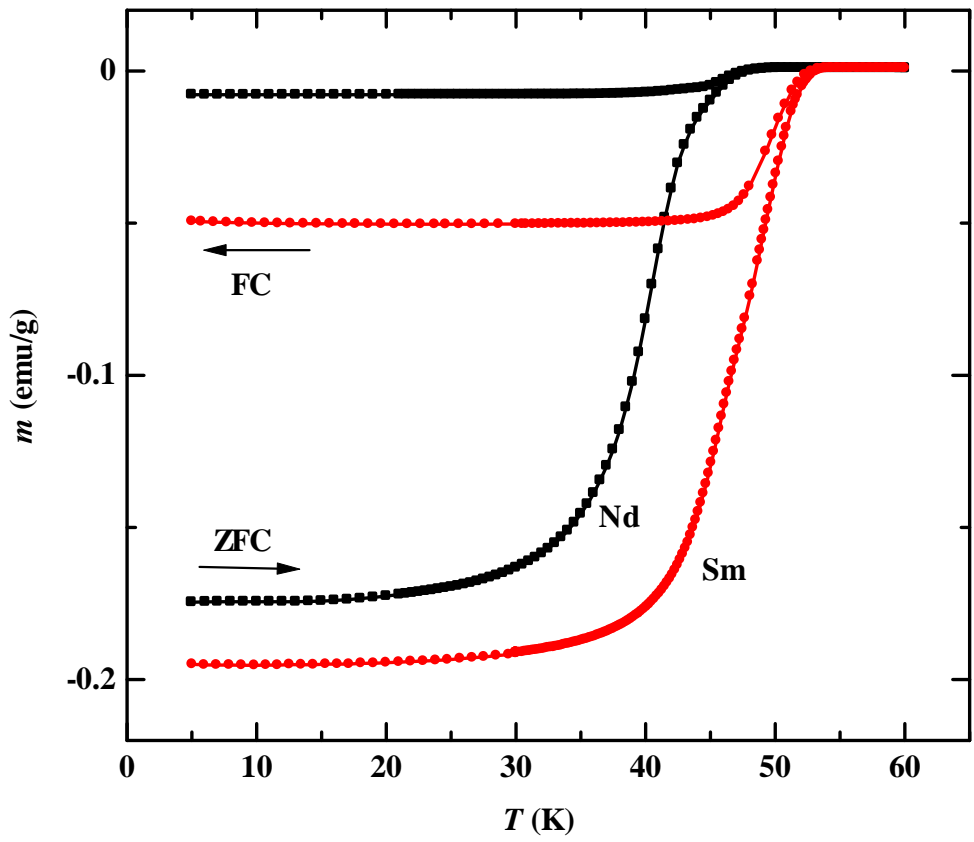

Figure 3

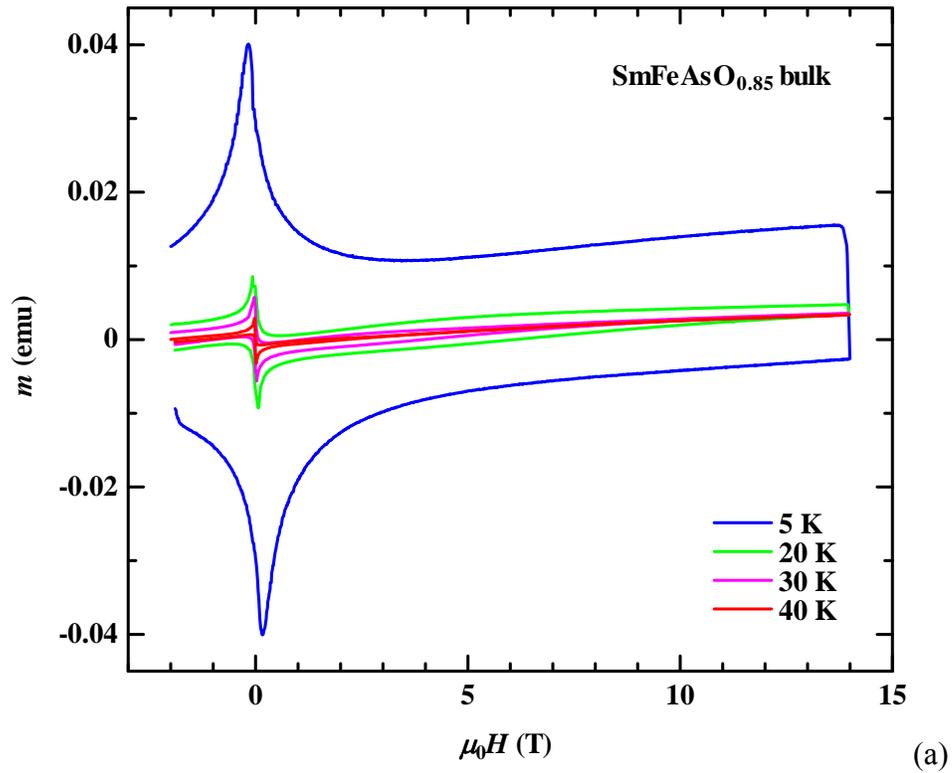

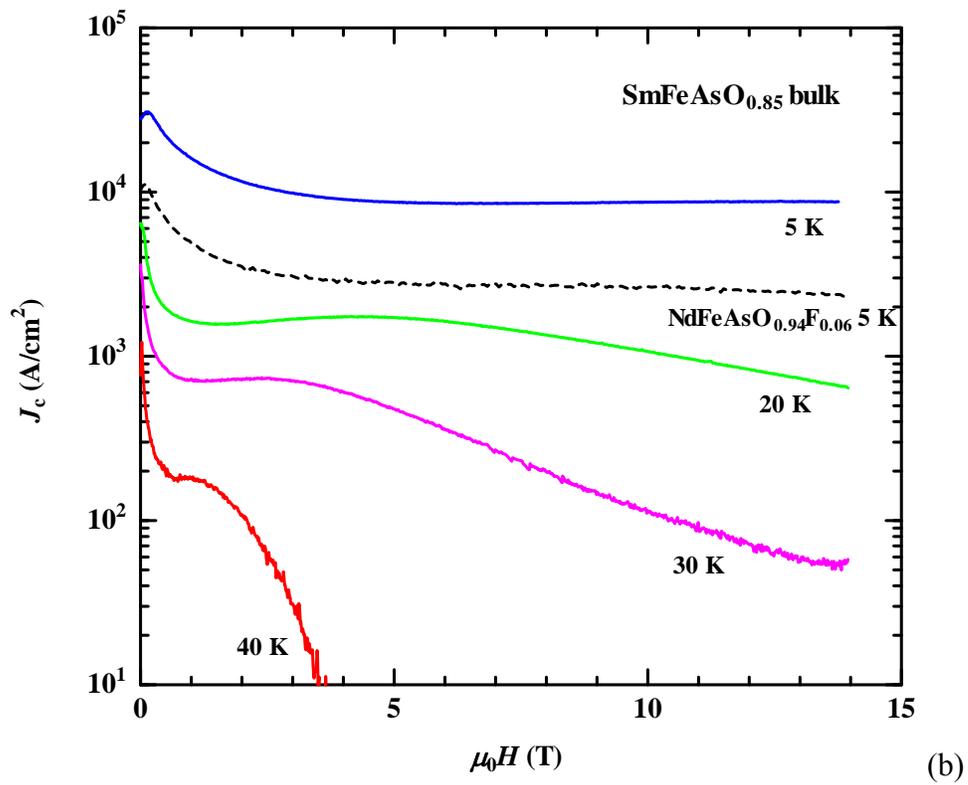

Figure 4

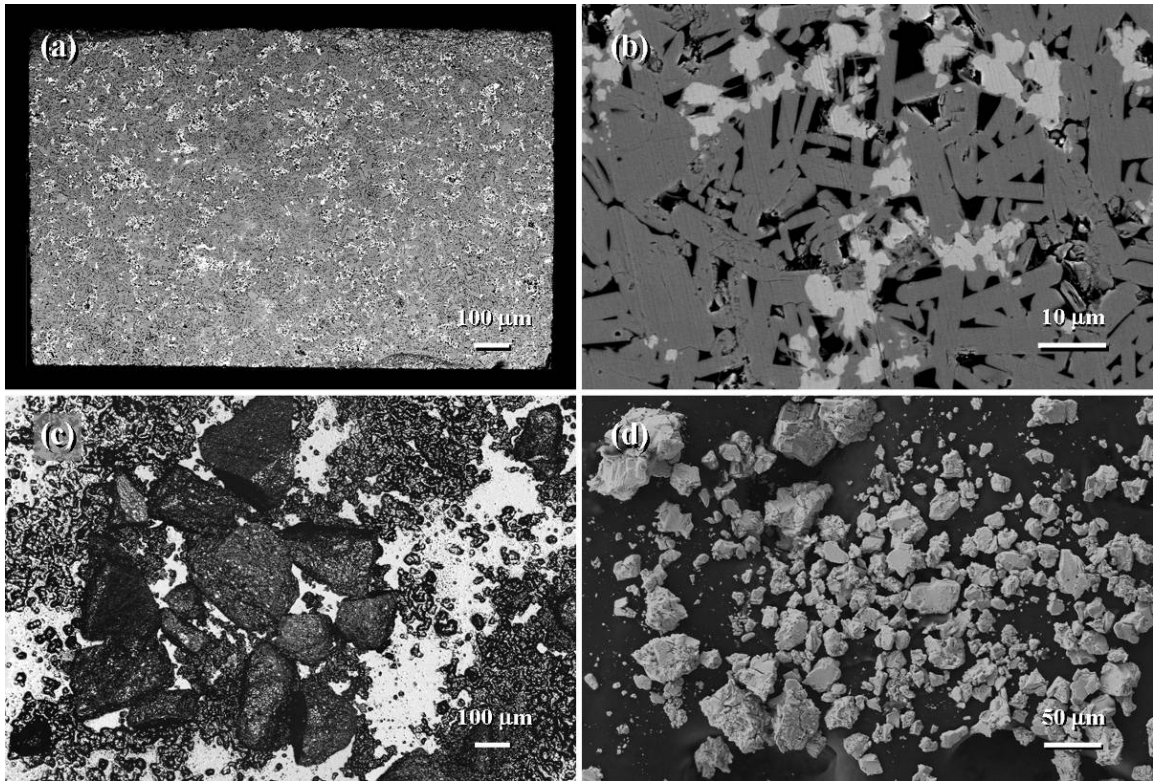

Figure 5

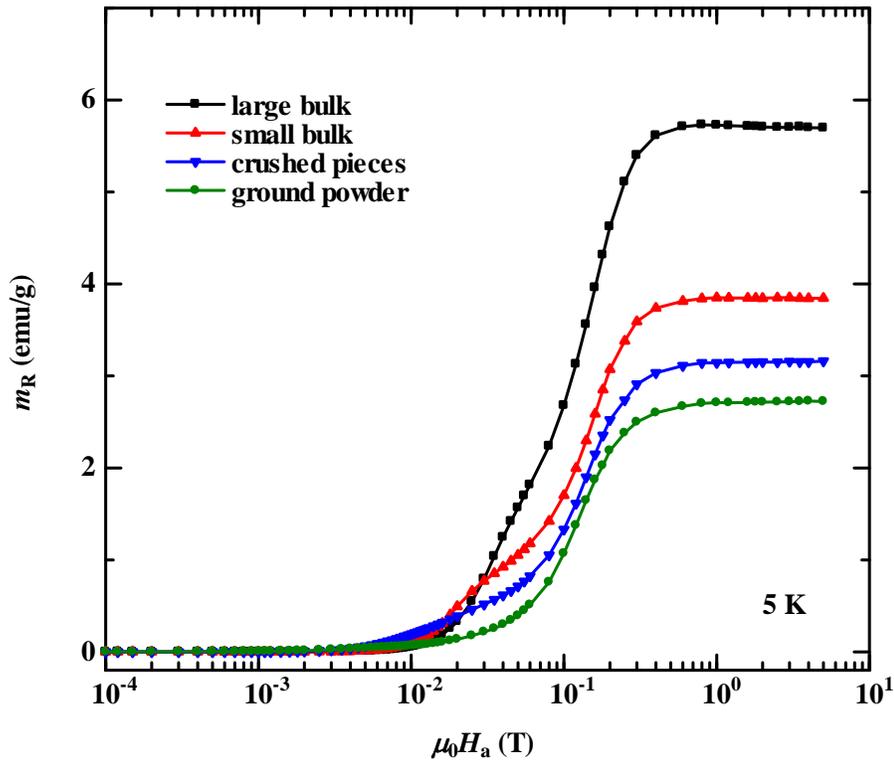

(a)

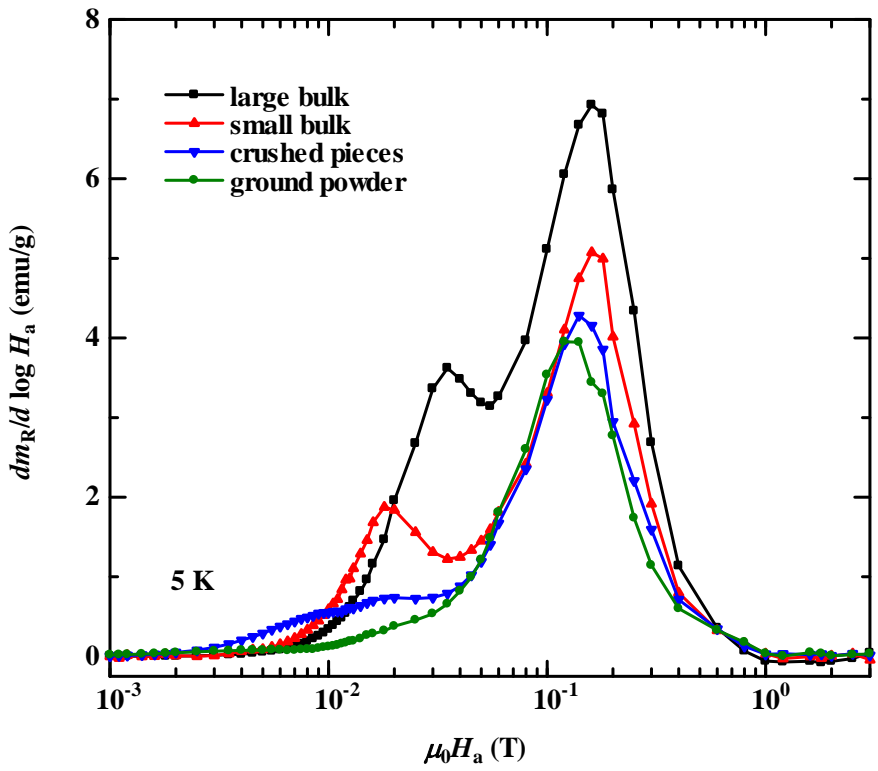

(b)

Figure 6

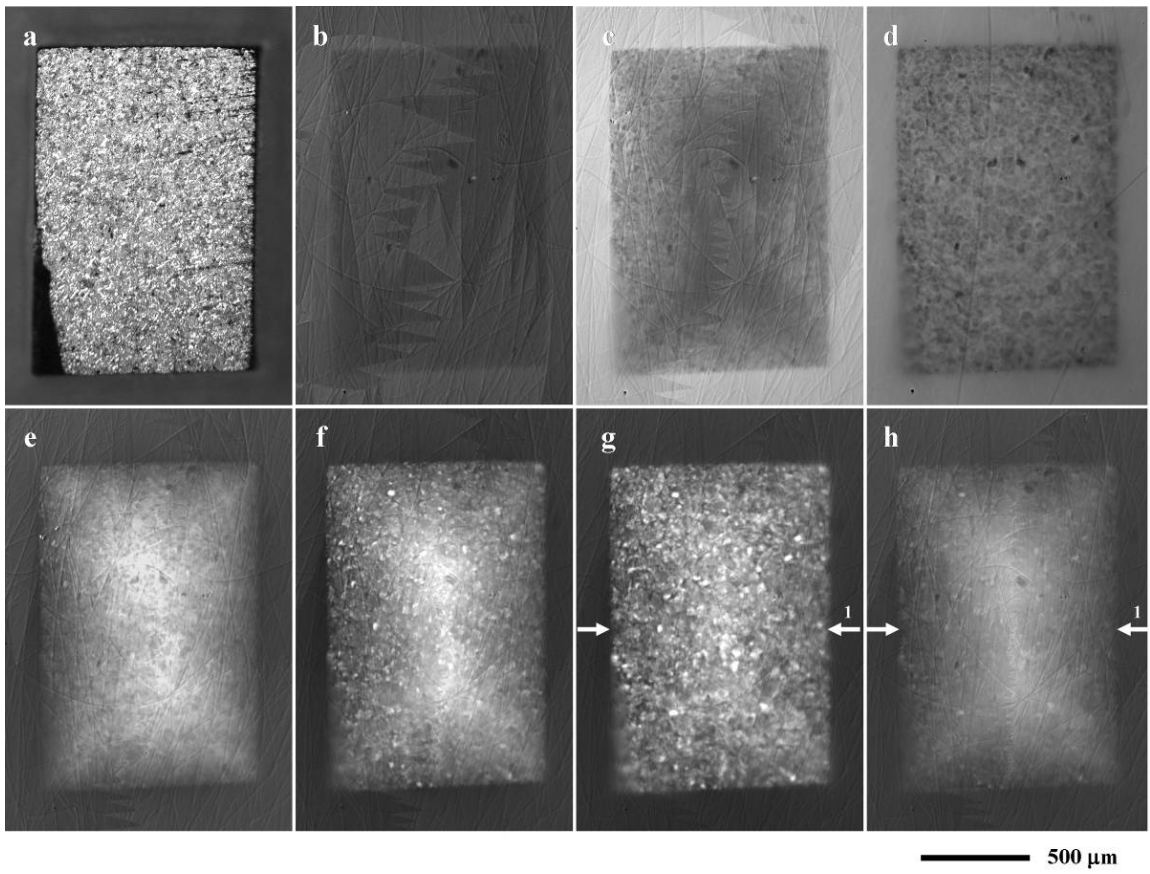

Figure 7(a,b)

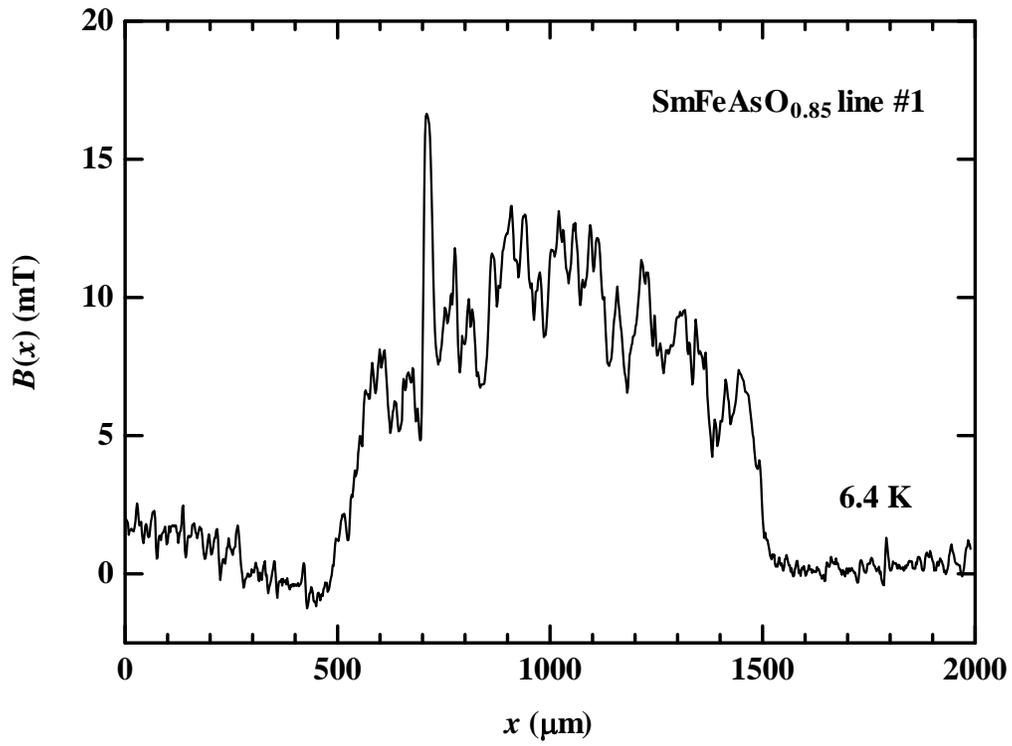

(a)

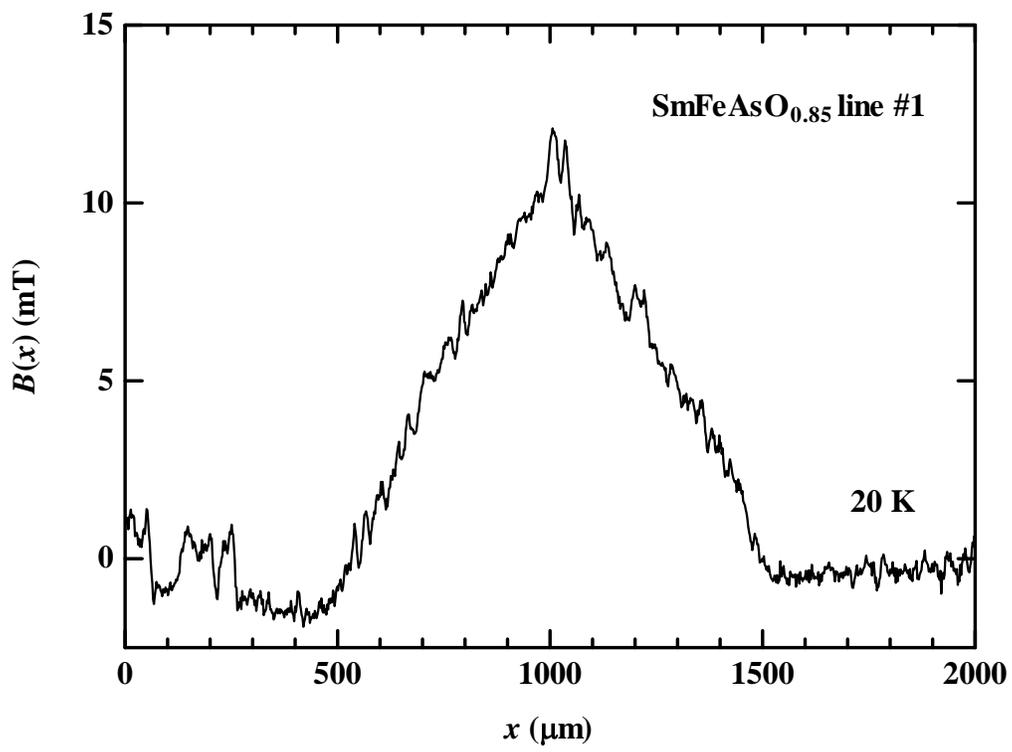

(b)

Figure 8

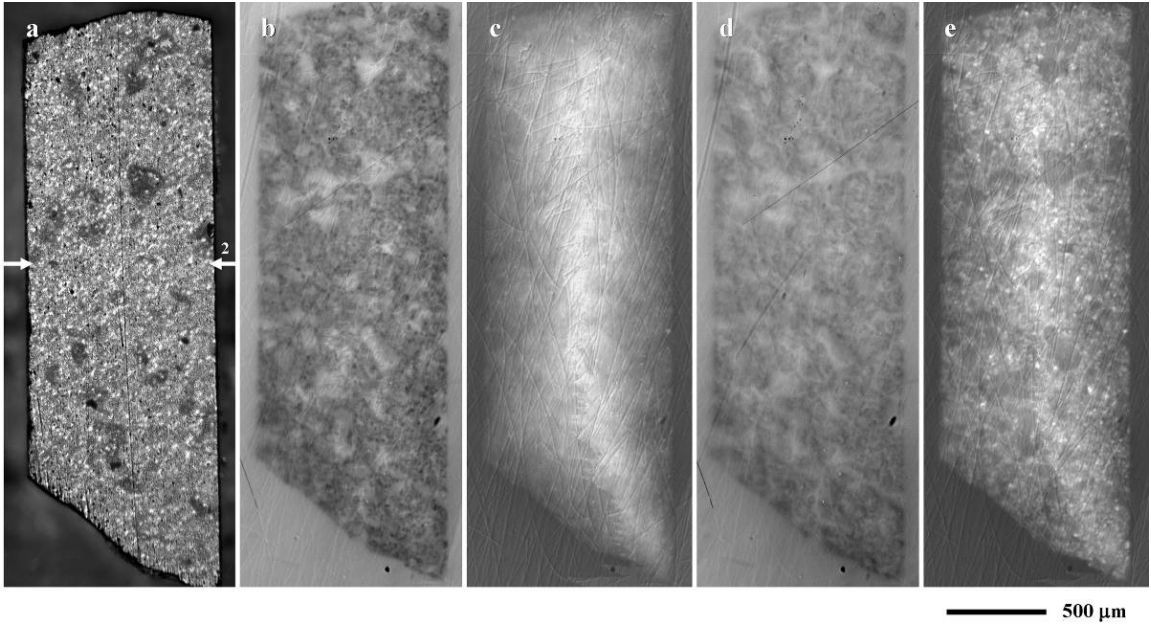

Figure 9(a,b)

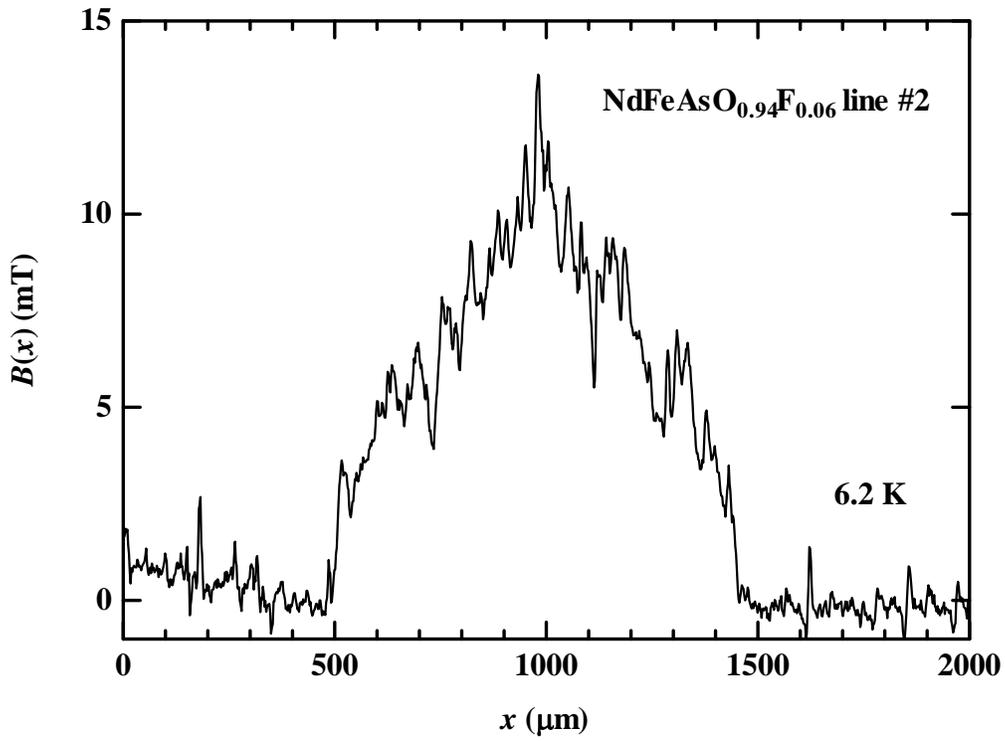

(a)

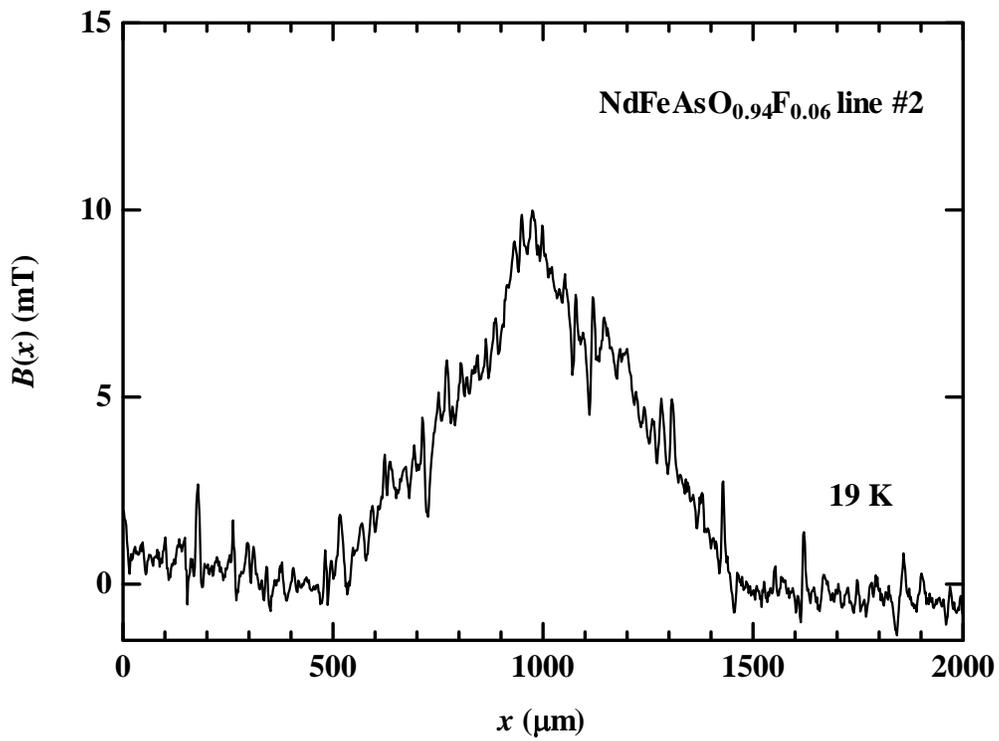

(b)

Figure 10

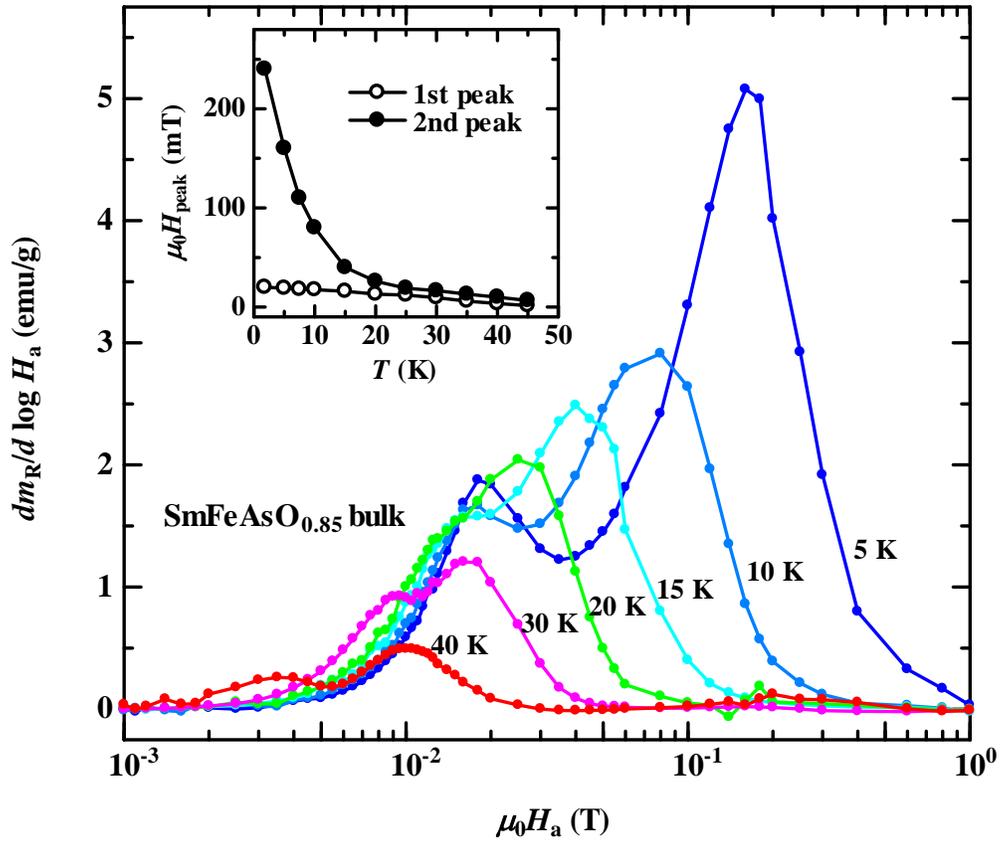

Figure 11

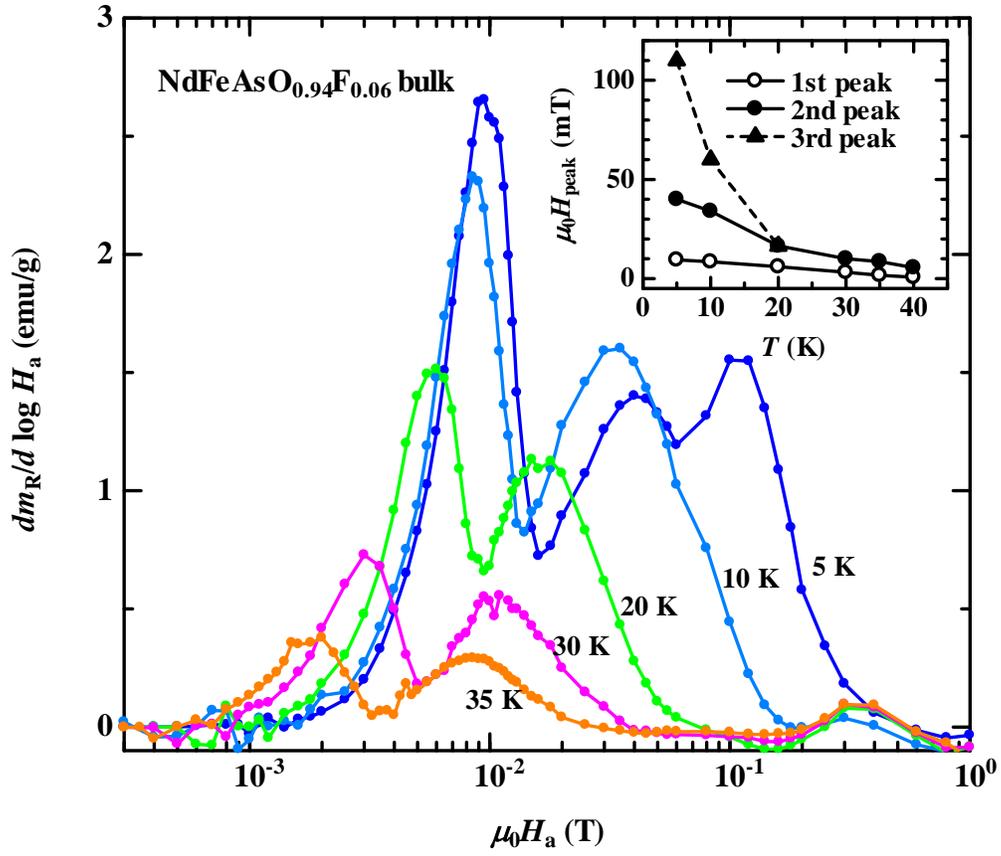

Figure 12

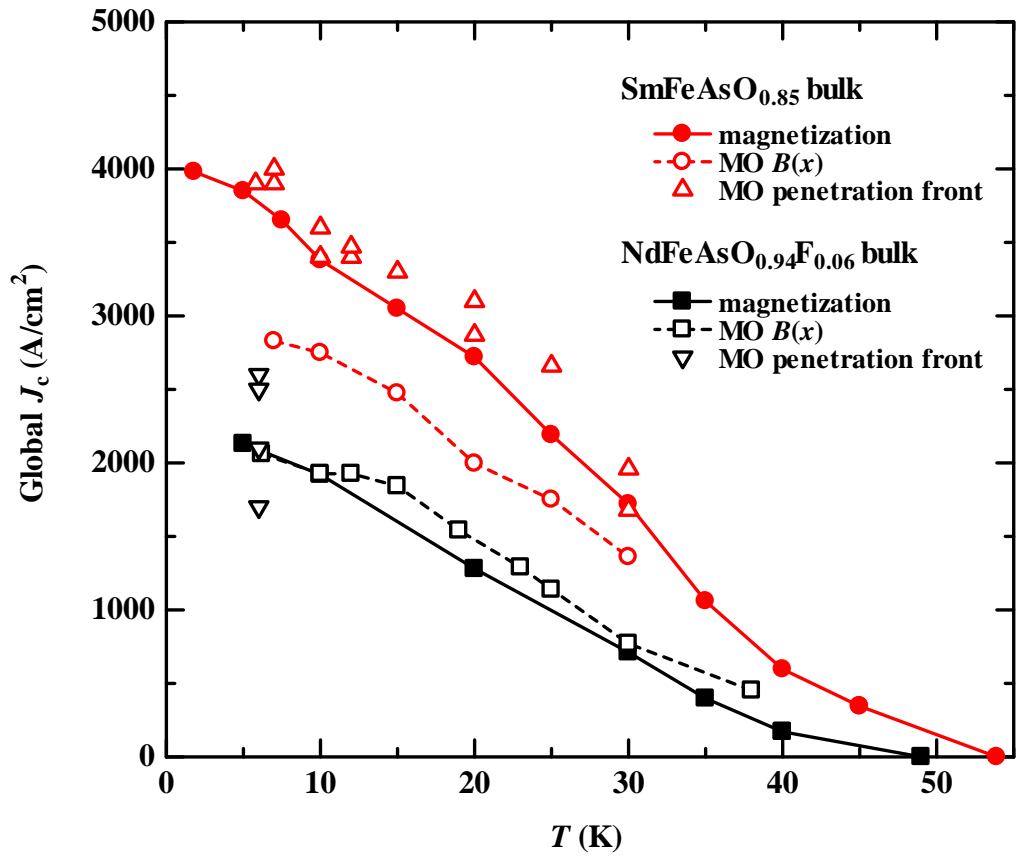

Figure 13

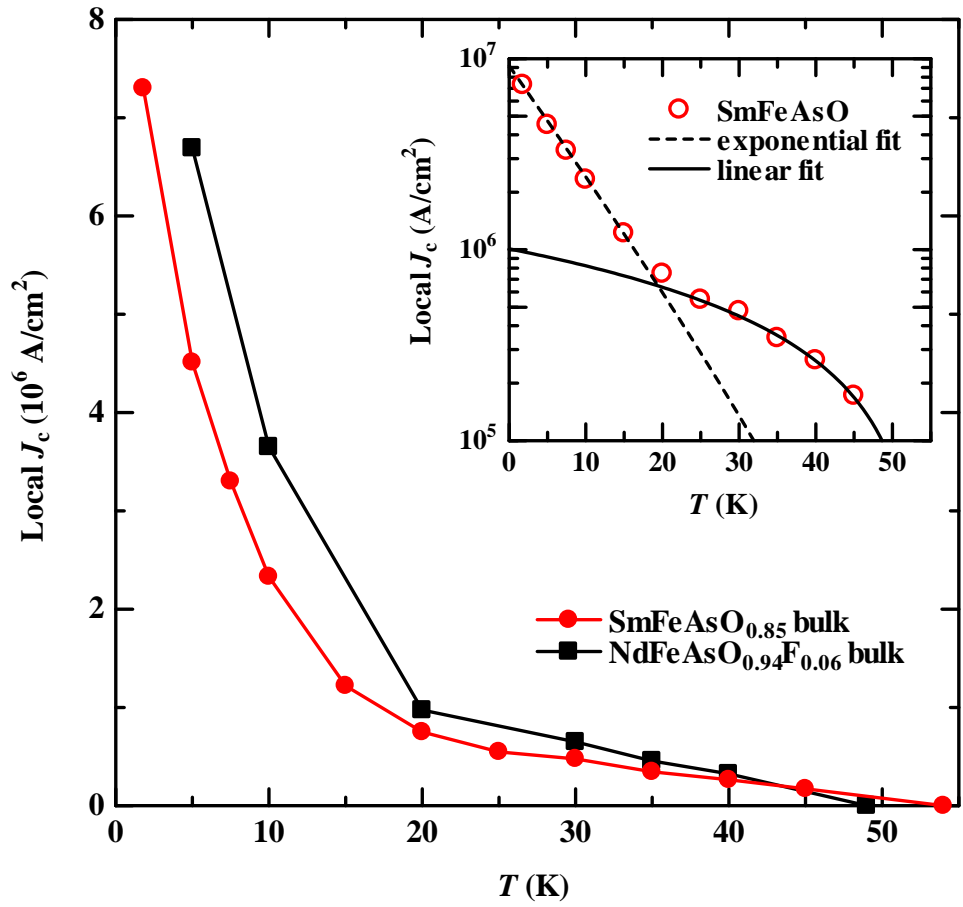

Figure 14

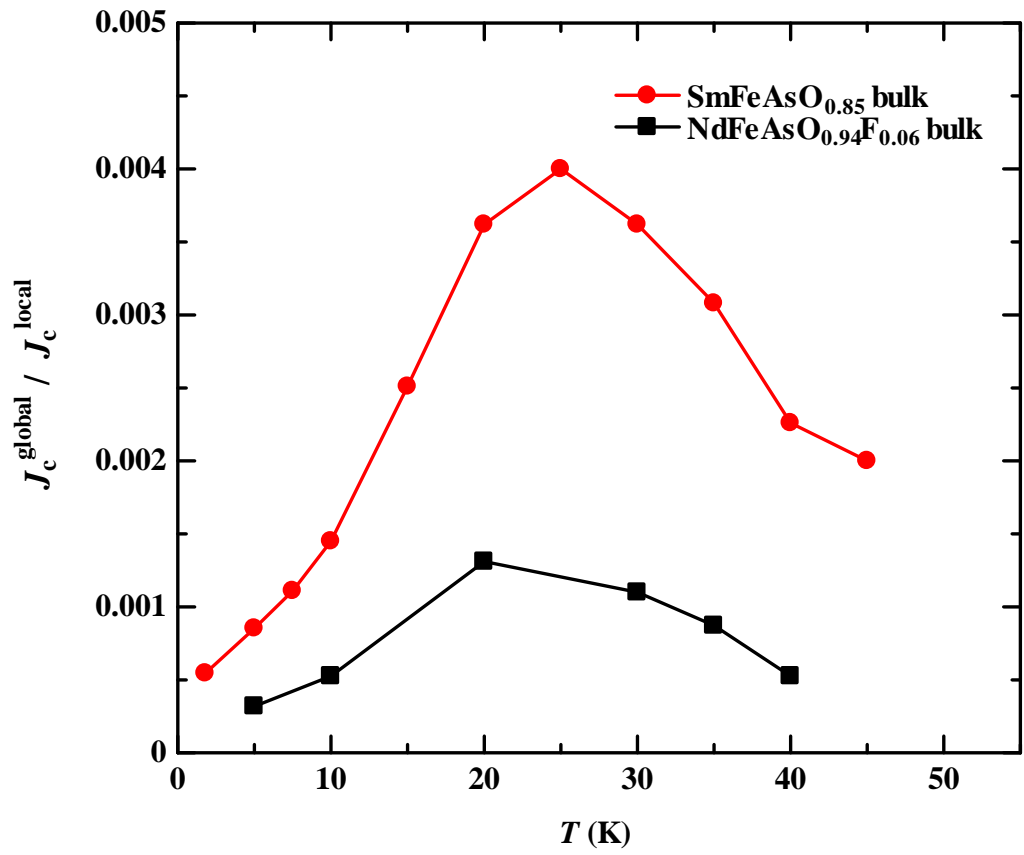